\documentclass[12pt]{article} 

\usepackage{tikz}
\usetikzlibrary{arrows,shapes,positioning}
\usetikzlibrary{decorations.markings}
\tikzstyle arrowstyle=[scale=1.6]
\tikzstyle directed=[postaction={decorate,decoration={markings,
    mark=at position .65 with {\arrow[arrowstyle]{stealth}}}}]
\tikzstyle reverse directed=[postaction={decorate,decoration={markings,
    mark=at position .65 with {\arrowreversed[arrowstyle]{stealth};}}}]
\usetikzlibrary{arrows}
\usepackage[hang,small,bf]{caption}
\usepackage{wrapfig}
\usepackage{caption}
\usepackage{subcaption}

\usepackage{color,epsfig,amsfonts,amssymb}

 \newcommand{\comma}{\;\; , \; \; }
\newcommand{\period}{\, \,  .}
\newcommand{\eq}{\; = \;}
\newcommand{\sep}{\;\; , \;\;}
\newcommand{\s}{, \,}
\newcommand{\negspce}{\! \! \! \! \! \! \! \! \!}
\newcommand{\spce}{\;\;\;\;\;\;}
\newcommand{\be}{\begin{equation}}
\newcommand{\bd}{\begin{displaymath}}
\newcommand{\ee}{\end{equation}}

\newcommand{\half}{{\textstyle \frac{1}{2}}}

\newcommand{\ed}{\end{displaymath}}
\newcommand{\ba}{\begin{eqnarray}}
\newcommand{\ea}{\end{eqnarray}}

\newcommand{\mybf}{\normalfont \bfseries}

\newcommand{\myit}{\normalfont \itshape}
\newcommand{\itt}{\normalfont \itshape}

\newcommand{\e}{{\mathrm{e}}}

\newcounter{storeeqn}

\title{Surface and corner free energies of the self-dual Potts model}

\author{ R.J. Baxter\\
{\protect \small  Mathematical
Sciences Institute}\\
{\protect \small  The Australian National University,
 Canberra, A.C.T. 0200, Australia  }
}

\date{\small 9 Nov 2016 }

\begin{document}

\definecolor{red1}{rgb}{0.8,0,0}

\definecolor{green1}{rgb}{0,0.5,0}

\definecolor{blue1}{rgb}{0,0,0.8}


\maketitle


\abstract{We consider the bulk, vertical surface, horizontal surface and
corner free energies  $f_b, f_s, f'_s, f_c$  of the anisotropic self-dual 
$Q$-state Potts model for $Q > 4$. $f_b$ was calculated in 
1973{\color{blue}\cite{RJB1973}}. For $Q<4$, $f_s, f'_s$ were calculated in 
1989{\color{blue}\cite{OB1989}}. Here we extend this last calculation to $Q>4$
and find agreement with the conjectures made in 2012 by 
Vernier and Jacobsen (VJ){\color{blue}\cite{VJ2012}} for the isotropic case.  
All these four free energies 
satisfy inversion and rotation relations. Together
with some plausible analyticity assumptions, these provide a less rigorous,
but much simpler, way of determining $f_b, f_s, f'_s$. They also imply
that $f_c$ is independent of the anisotropy, being a function only of $Q$, 
in which respect they resemble the order parameters of the associated six-vertex model.
Hence VJ's conjecture for $f_c$ should apply to the full anisotropic model.}



\renewcommand{\theequation}{\arabic{section}.\arabic{equation}}

 
\tableofcontents

\section{Introduction}

Vernier and Jacobsen{\color{blue}\cite{VJ2012}} considered a number of two-dimensional 
lattice models in statistical mechanics for which the bulk free energies have been 
calculated exactly and conjectured their surface and corner free energies. They 
considered only the rotation-invariant (isotropic) cases of these models, when the surface
 free energies are the same for the vertical and horizontal surfaces.

For some of these models the surface free energies  have been, or can readily be,
calculated exactly, and this can be done for the more general non-rotation-invariant cases. 
For the case of the square-lattice self-dual Potts model, Vernier and Jacobsen
commented that it seemed likely that the surface free energy had been calculated. It 
seems that this has not yet been reported in the literature for the case in which they were 
interested. That omission is repaired here for the general anisotropic case.

We also present arguments that Vernier and Jacobsen's{\color{blue}\cite{VJ2012}} 
conjecture for the corner free energy should apply to the anisotropic case.


Consider the self-dual $Q$-state Potts model on the square lattice, which is 
equivalent to an homogeneous six-vertex model.{\color{blue}\cite[\S 12.5]{book}} 
Owczarek and Baxter{\color{blue}\cite{OB1989}} showed that for this model an extended 
Bethe  ansatz worked for a lattice of $N$ columns with free (rather than cylindrical) 
boundary conditions.
They wrote down the resulting ``Bethe equations" for the eigenvalues of the 
row-to-row transfer matrix  $T$. They were
interested in the critical case, which occurs when the number of states $Q$ is not greater 
than 4, and solved the equations for $N$ large to obtain the bulk and 
surface free energies.

Vernier and Jacobsen{\color{blue}\cite[\S 3.2.1]{VJ2012}} instead considered the case 
$Q > 4$, when the model is at a first-order transition point.
Here we solve the Bethe equations for this case. We obtain the surface
free energies  $f_s$ and $f'_s$ (as well as the bulk free energy $f_b$) and verify the 
correctness of Vernier and  Jacobsen's conjectures for the rotation-invariant case.

We also show that the four free energies all satisfy ``inversion" and ``rotation" relations,
and that if we assume certain plausible analyticity properties, then these are sufficient
to determine the bulk and surface free energies, and to show that the corner free energy
is independent of the anisotropy of the model, depending only on $Q$. The results
of this method of course agree with those of the more rigorous Bethe ansatz 
calculations.
 
The self-dual Potts model contains two free parameters  $Q, K_1$, or equivalently the 
$q, w$ defined by (\ref{weights}), (\ref{xxx}), (\ref{defx}), (\ref{defqw}).\footnote{{From}
 these equations, $Q = q+2+q^{-1}$.}
Our Bethe ansatz method is not sufficient to calculate the corner free energy $f_c$ ,
but the inversion relation method implies that it is independent of $K_1$ or $w$,
 depending only on $Q$ or $q$. We  also comment in section \ref{sec4} that we have 
 performed direct  numerical calculations on 
 finite lattices to obtain the first 10 coefficients in a series expansion in powers of $q$ as
  functions of $s = w^2/q^{1/2}$.  (Each coefficient is a finite Laurent polynomial in $s$.)
  We  find agreement (as expected) with Vernier and 
 Jacobsen's,{\color{blue}\cite[\S 3.2.1]{VJ2012}} conjecture for the isotropic 
 case,\footnote{$q$ herein is  $q_{VJ}^2$, where $q_{VJ}$ is the $q$ of Vernier 
 and Jacobsen, and all the free energies are negated.} which is when
 $w = q^{1/4}$ and $s=1$. 
 
For the corner free energy $f_c$, we  also observe that all the 10 coefficients are 
{\em independent } of $s$, which agrees with  the inversion relation result that $f_c$  
is  a function only  of $q$.
 
For this model, therefore, $f_c$ resembles the order parameters $M_0$ and $P_0$
of the  associated six-vertex model,{\color{blue}\cite[eqn. 8.10.9]{book}}, in that it
depends only on $Q$ or $q$.
 
 We have found corresponding behaviour for the square-lattice Ising 
 model.{\color{blue}\cite{Ising}} For both models, this means that the corner free energy
 is a function only of the order parameter. Possibly this property applies more generally.



\section{The square-lattice Potts model}
\setcounter{equation}{0}

We consider the $Q$-state Potts model on a square lattice $\cal L$ of $M$ rows and 
$N$ columns, as shown in Fig. {\ref{sqlattice1}}. On each site $i$ there is a "spin" $\sigma_i$ that 
takes the values $1, 2, \ldots ,Q$. Spins at horizontally adjacent sites $i,  j$ 
interact with dimensionless energy $-K_1 \delta (\sigma_i, \sigma_j)$, and
those on vertically adjacent sites with energy $-K_2 \delta (\sigma_k, \sigma_m)$.

\setlength{\unitlength}{1.2pt}
\begin{figure}[hbt]
\begin{picture}(420,160) (0,0)

\put (174,7) { $K_1$ }
\put (144,-12) { $i$ }
\put (204,-12) { $j$ }

\put (253,90) { $K_2$ }
\put (275,59) { $k$ }
\put (273,119) { $m$ }

\put (85,-13) { $1$ }
\put (265,-13) { $N$ }

\put (73,-1) { $1$ }
\put (72,119) { $M$ }

{\color{blue} 
\multiput(91.4,0.5)(60,0){4}{\circle{7}}
\multiput(91.4,60.5)(60,0){4}{\circle{7}}
\multiput(91.4,120.5)(60,0){4}{\circle{7}}

\multiput(95.2,00)(5,0){11}{\bf .}
\multiput(155.2,00)(5,0){11}{\bf .}
\multiput(215.2,00)(5,0){11}{\bf .}

\multiput(95.2,60)(5,0){11}{\bf .}
\multiput(155.2,60)(5,0){11}{\bf .}
\multiput(215.2,60)(5,0){11}{\bf .}

\multiput(95.2,120)(5,0){11}{\bf .}
\multiput(155.2,120)(5,0){11}{\bf .}
\multiput(215.2,120)(5,0){11}{\bf .}

\put (304, 25) {\LARGE  $\cal L$ }

\thinlines

\multiput(90,4.6)(0,5){11}{\bf .}
\multiput(90,64.6)(0,5){11}{\bf .}

\multiput(150,4.6)(0,5){11}{\bf .}
\multiput(150,64.6)(0,5){11}{\bf .}

\multiput(210,4.6)(0,5){11}{\bf .}
\multiput(210,64.6)(0,5){11}{\bf .}

\multiput(270,4.6)(0,5){11}{\bf .}
\multiput(270,64.6)(0,5){11}{\bf .}}

 \end{picture}
\vspace{1.0cm}

  \caption{ The square lattice $\cal L$ (of 3 rows and 4 columns), indicating the horizontal 
  and vertical interaction coefficients  $K_1, K_2$.}

 \label{sqlattice1}
\end{figure}
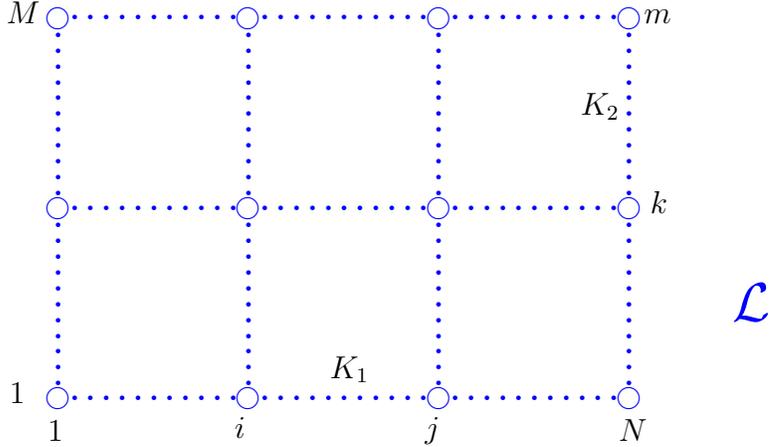

The partition function is 

\be \label{Pottspartnfn}
Z_P \eq \sum_{\bf \sigma }
\exp \left[ K_1 \sum \delta (\sigma_i, \sigma_j) + 
K_2 \sum \delta (\sigma_k, \sigma_m) \right] \comma \ee
where the first inner sum is over all horizontal edges $(i,j)$ and the second over all
vertical edges $(k,m)$. The outer sum is over all $Q^{MN}$ values of all the spins.

We expect that when $M, N$ are large, 
\be  \label{freeenergies}
\log Z_P \eq -MN f_b - M f_s - N f'_s - f_c + O(\e^{-\delta M},\e^{-\delta' N})
\comma \ee
where $f_b, f_s, f'_s, f_c $ are the dimensionless bulk, vertical surface, horizontal surface
 and corner free energies,
and $\delta, \delta'$ are positive numbers.

\setlength{\unitlength}{1.2pt}
\begin{figure}[hbt]
\begin{picture}(420,160) (0,0)
\put (60,120) {\line(1,1) {30}}
\put (60,0) {\line(1,1) {150}}
\put (60,60) {\line(1,1) {90}}
\put (90,-30) {\line(1,1) {180}}
\put (150,-30) {\line(1,1) {150}}
\put (210,-30) {\line(1,1) {90}}
\put (270,-30) {\line(1,1) {30}}

\put (60,00) {\line(1,-1) {30}}
\put (60,120) {\line(1,-1) {150}}
\put (60,60) {\line(1,-1) {90}}
\put (90,150) {\line(1,-1) {180}}
\put (150,150) {\line(1,-1) {150}}
\put (210,150) {\line(1,-1) {90}}
\put (270,150) {\line(1,-1) {30}}

\put (320, 30) {\LARGE  $\cal L'$ }

{\color{blue} 
\multiput(90,0.5)(60,0){4}{\circle{7}}
\multiput(90,60.5)(60,0){4}{\circle{7}}
\multiput(90,120.5)(60,0){4}{\circle{7}}

\multiput(93.5,00)(5,0){11}{\bf .}
\multiput(153.5,00)(5,0){11}{\bf .}
\multiput(213.5,00)(5,0){11}{\bf .}

\multiput(93.5,60)(5,0){11}{\bf .}
\multiput(153.5,60)(5,0){11}{\bf .}
\multiput(213.5,60)(5,0){11}{\bf .}

\multiput(93.5,120)(5,0){11}{\bf .}
\multiput(153.5,120)(5,0){11}{\bf .}
\multiput(213.5,120)(5,0){11}{\bf .}

\thinlines

\multiput(88.6,4.6)(0,5){11}{\bf .}
\multiput(88.6,64.6)(0,5){11}{\bf .}

\multiput(148.6,4.6)(0,5){11}{\bf .}
\multiput(148.6,64.6)(0,5){11}{\bf .}

\multiput(208.6,4.6)(0,5){11}{\bf .}
\multiput(208.6,64.6)(0,5){11}{\bf .}

\multiput(268.6,4.6)(0,5){11}{\bf .}
\multiput(268.6,64.6)(0,5){11}{\bf .}}

\multiput(90,-30)(60,0){4}{\circle*{7}}
\multiput(60,0)(60,0){5}{\circle*{7}}
\multiput(60,60)(60,0){5}{\circle*{7}}
\multiput(60,120)(60,0){5}{\circle*{7}}
\multiput(90,30)(60,0){4}{\circle*{7}}
\multiput(90,90)(60,0){4}{\circle*{7}}
\multiput(90,150)(60,0){4}{\circle*{7}}

\put (25,-30) {$1$}
\put (25,0) {$2$}
\put (25,120) {$2M$}
\put (18,150) {$2M \! + \! 1$}

\put (87,-50) {$1$}
\put (147,-50) {$2$}
\put (265,-50) {$N$}

 \end{picture}
\vspace{2.0cm}

\caption{\footnotesize The square lattice $\cal L$  of dotted lines and 
circles,  and its medial lattice $\cal L'$ of full circles and lines .}

 \label{sqlattice2}
\end{figure}
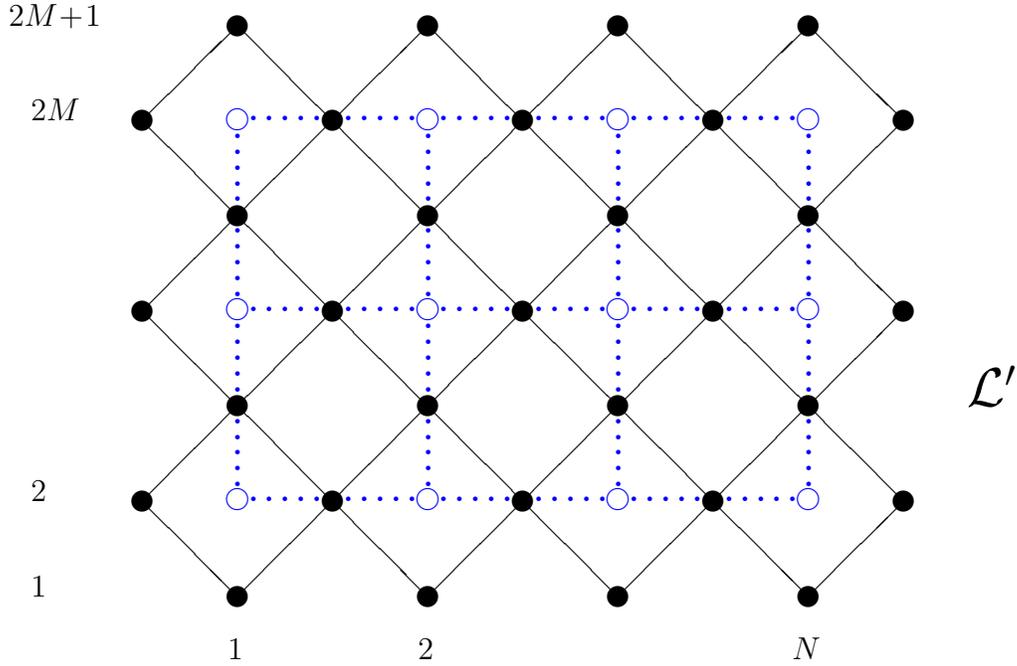

We show in {\color{blue}\cite[\S 12.5]{book}} that this model is equivalent to a 
six-vertex  model on the lattice $\cal L'$ of Fig. \ref{sqlattice2}, i.e the lattice of 
solid lines and circles therein. On this lattice we place an arrow on each edge subject 
to the rule that at  each site or vertex there must be as many arrows pointing in  as
there are  pointing out. There are six such configurations of arrows at an internal  vertex, as 
shown  in Fig. {\ref{sixvertices}}.


The lattice $\cal L'$ has $2M+1$ rows, even-numbered rows having $N+1$ 
vertices,  and odd-numbered ones having $N$ vertices.
 Between two successive rows there are $2N$ diagonal 
edges, on which one places arrows. Each of the $M$ even-numbered rows  has
$N-1$ internal vertices, with  weights
\be \label{oddrows} \omega_1, \ldots , \omega_6 \eq 1 \s   \s  x_1 \s  x_1 \s  
1+x_1 \e^{\lambda} \s 1+x_1 \e^{-\lambda} \comma \ee
and each of the $M-1$ odd-numbered rows $3, 5, \ldots, 2M-1$ has
$N$ internal vertices with weights
\be  \label{evenrows} \omega_1, \ldots , \omega_6 \eq x_2 \s x_2 \s  1 \s  1 \s  
x_2+\e^{\lambda} \s   x_2+ \e^{-\lambda} \comma \ee
where 
\be \label{weights}
Q^{1/2} \eq 2 \, \cosh \lambda \sep x_1= (\e^{K_1}-1)/Q^{1/2}  \sep x_2 = 
(\e^{K_2}-1)/Q^{1/2}  \period  \ee
The vertices on the boundaries of $\cal L'$ only have two edges joining them and must 
have one arrow in and one arrow out. The weights of the possible configurations are
indicated in Fig. \ref{bdyweights}.

The partition function of this six-vertex model is
\be Z_{6V} \eq \sum_C  \;  \; {\displaystyle \prod }_i  \; w_i \comma \ee
where the sum is over all allowed configurations $C$ of arrows on the edges of
$\cal L'$ and for each configuration the product is over all vertices $i$ of the 
corresponding weights $w_i$ (including the boundary vertices).

If $\cal L'$ were wound on a torus (which is {\em not} the case considered in this 
paper),  we could interchange the two types of rows without
affecting the partition function. This is equivalent to replacing $x_1, x_2$ by
$x_1^{*} = 1/x_2, x_2^{*} = 1/x_1$ and multiplying $Z_P$ by $(x_1/x_2)^{MN}$, and
 to replacing $K_1, K_2$ by their ``duals"  $K_1^{*}, K_2^{*}$, where
 \be \label{dualKK}
  \exp( {K_1^{*}} ) =  \frac{\e^{K_2}+Q-1}{\e^{K_2} -1} \sep 
  \exp( {K_2^{*}} ) =  \frac{\e^{K_1}+Q-1}{\e^{K_1} -1}  \period \ee

The partition function $Z$ of the Potts model, as defined in (\ref{Pottspartnfn}), is 
related exactly to $Z_{6V}$ by
\be \label{Pottsand 6V}
Z_P \eq Q^{MN/2} \, Z_{6V}  \ee



\vspace{0.6cm}

\setlength{\unitlength}{1pt}
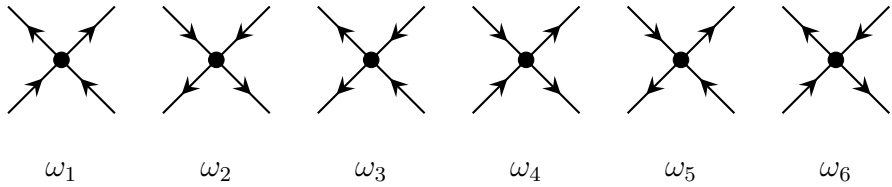
\begin{figure}[hbt]
\begin{picture}(420,90) (-50,0)
\begin{tikzpicture}[scale=0.71]


\draw[black,fill=black] (1.0,-2) circle (0.8ex);
\draw[black,fill=black] (3.9,-2) circle (0.8ex);
\draw[black,fill=black] (6.8,-2) circle (0.8ex);
\draw[black,fill=black] (9.7,-2) circle (0.8ex);
\draw[black,fill=black] (12.6,-2) circle (0.8ex);
\draw[black,fill=black] (15.5,-2) circle (0.8ex);

 \node at (1.0,-4.1) { $\omega_1$};
\node at (3.9,-4.1) { $\omega_2$};
 \node at (6.8,-4.1) { $\omega_3$};
 \node at (9.7,-4.1) { $\omega_4$};
 \node at (12.6,-4.1) { $\omega_5$};
 \node at (15.5,-4.1) { $\omega_6$};


 \coordinate (Q) at (1,-2) ;
  \coordinate (R1) at (2,-1) ;
  \coordinate (R2) at (0,-3) ;
 \coordinate (R3) at (2,-3) ;
 \coordinate (R4) at (0,-1) ;
 
  \coordinate (Q2) at (3.9,-2) ;
  \coordinate (S1) at (4.9,-1) ;
  \coordinate (S2) at (2.9,-3) ;
 \coordinate (S3) at (4.9,-3) ;
 \coordinate (S4) at (2.9,-1) ;
 
  \coordinate (Q3) at (6.8,-2) ;
  \coordinate (T1) at (7.8,-1) ;
  \coordinate (T2) at (5.8,-3) ;
 \coordinate (T3) at (7.8,-3) ;
 \coordinate (T4) at (5.8,-1) ;
 
  \coordinate (Q4) at (9.7,-2) ;
  \coordinate (U1) at (10.7,-1) ;
  \coordinate (U2) at (8.7,-3) ;
 \coordinate (U3) at (10.7,-3) ;
 \coordinate (U4) at (8.7,-1) ;
 
   \coordinate (Q5) at (12.6,-2) ;
  \coordinate (V1) at (13.6,-1) ;
  \coordinate (V2) at (11.6,-3) ;
 \coordinate (V3) at (13.6,-3) ;
 \coordinate (V4) at (11.6,-1) ;
 
  \coordinate (Q6) at (15.5,-2) ;
  \coordinate (W1) at (16.5,-1) ;
  \coordinate (W2) at (14.5,-3) ;
 \coordinate (W3) at (16.5,-3) ;
 \coordinate (W4) at (14.5,-1) ;

 \draw[black,directed,thick] (Q) -- (R4);
\draw[black,directed,thick] (R3) -- (Q);
\draw[black,directed,thick] (Q) -- (R1);
\draw[black,directed,thick] (R2) -- (Q);

 \draw[black,directed,thick] (S4) -- (Q2);
\draw[black,directed,thick] (Q2) -- (S3);
\draw[black,directed,thick] (S1) -- (Q2);
\draw[black,directed,thick] (Q2) -- (S2);

\draw[black,directed,thick] (Q3)--(T4);
\draw[black,directed,thick] (T3) -- (Q3);
\draw[black,directed,thick] (T1) -- (Q3);
\draw[black,directed,thick] (Q3) -- (T2);

\draw[black,directed,thick] (U4) --(Q4);
\draw[black,directed,thick] (Q4)--(U3);
\draw[black,directed,thick] (Q4)--(U1);
\draw[black,directed,thick]  (U2)--(Q4);

\draw[black,directed,thick] (V4) --(Q5);
\draw[black,directed,thick] (V3)--(Q5);
\draw[black,directed,thick] (Q5)--(V1);
\draw[black,directed,thick]  (Q5)--(V2);

\draw[black,directed,thick] (Q6)--(W4);
\draw[black,directed,thick] (Q6)--(W3);
\draw[black,directed,thick] (W1)--(Q6);
\draw[black,directed,thick]  (W2)--(Q6);
\end{tikzpicture}
\end{picture}

\vspace{0.0cm}
 \caption{\footnotesize The six vertices, with weights $\omega_1, \ldots ,\omega_6$.}
 \label{sixvertices}
\end{figure}


\vspace{0.3cm}


\setlength{\unitlength}{1pt}
\begin{figure}[hbt]
\begin{picture}(420,70) (-52,10)
\begin{tikzpicture}[scale=0.51]


\draw[black,fill=black] (1.0,-2) circle (0.8ex);
\draw[black,fill=black] (3.9,-2) circle (0.8ex);
\draw[black,fill=black] (7.2,-1.4) circle (0.8ex);
\draw[black,fill=black] (10.1,-1.4) circle (0.8ex);
\draw[black,fill=black] (13,-1.7) circle (0.8ex);
\draw[black,fill=black] (15.6,-1.7) circle (0.8ex);
\draw[black,fill=black] (19.7,-1.7) circle (0.8ex);
\draw[black,fill=black] (22.3,-1.7) circle (0.8ex);

 \node at (1.2,-4.1) { $\e^{\lambda/2}$};
\node at (4,-4.1) { $\e^{-\lambda/2}$};
\node at (7.2,-4.1) { $\e^{\lambda/2}$};
 \node at (10.1,-4.1) { $\e^{-\lambda/2}$};
 \node at (13.4,-4.1) {  \footnotesize$1$};
 \node at (16,-4.1) { \footnotesize $1$};
 \node at (19.5,-4.1) {  \footnotesize$1$};
 \node at (22.1,-4.1) { \footnotesize $1$};


 \coordinate (Q) at (1,-2) ;
  \coordinate (R1) at (2,-1) ;
 \coordinate (R4) at (0,-1) ;
 
 \coordinate (Q2) at (3.9,-2) ;
  \coordinate (S1) at (4.9,-1) ;
 \coordinate (S4) at (2.9,-1) ;

  \coordinate (Q3) at (7.2,-1.4) ;
  \coordinate (T2) at (6.2,-2.4) ;
 \coordinate (T3) at (8.2,-2.4) ;
 
  \coordinate (Q4) at (10.1,-1.4) ;
    \coordinate (U2) at (9.1,-2.4) ;
 \coordinate (U3) at (11.1,-2.4) ;

   \coordinate (Q5) at (13,-1.7) ;
  \coordinate (V1) at (14,-0.7) ;
  \coordinate (V3) at (14,-2.7) ;
  
  \coordinate (Q6) at (15.6,-1.7) ;
  \coordinate (W1) at (16.6,-0.7) ;
 \coordinate (W3) at (16.6,-2.7) ;

  \coordinate (Q7) at (19.7,-1.7) ;
  \coordinate (X2) at (18.7,-2.7) ;
 \coordinate (X4) at (18.7,-0.7) ;
 
  \coordinate (Q8) at (22.3,-1.7) ;
  \coordinate (Y2) at (21.3,-2.7) ;
 \coordinate (Y4) at (21.3,-0.7) ;

 \draw[black,directed,thick] (R4) -- (Q);
\draw[black,directed,thick] (Q) -- (R1);

 \draw[black,directed,thick] (Q2) -- (S4);
\draw[black,directed,thick] (S1) -- (Q2);

\draw[black,directed,thick] (T3) -- (Q3);
\draw[black,directed,thick] (Q3) -- (T2);

\draw[black,directed,thick] (Q4)--(U3);
\draw[black,directed,thick]  (U2)--(Q4);

\draw[black,directed,thick] (V3)--(Q5);
\draw[black,directed,thick] (Q5)--(V1);

\draw[black,directed,thick] (Q6)--(W3);
\draw[black,directed,thick] (W1)--(Q6);

\draw[black,directed,thick] (V3)--(Q5);
\draw[black,directed,thick] (Q5)--(V1);

\draw[black,directed,thick] (Q6)--(W3);
\draw[black,directed,thick] (W1)--(Q6);

\draw[black,directed,thick] (X4) --(Q7);
\draw[black,directed,thick]  (Q7)--(X2);

\draw[black,directed,thick] (Q8)--(Y4);
\draw[black,directed,thick]  (Y2)--(Q8);

\end{tikzpicture}
\end{picture}

\vspace{0.5cm}
\caption{\footnotesize The boundary weights.}
 \label{bdyweights}
\end{figure}
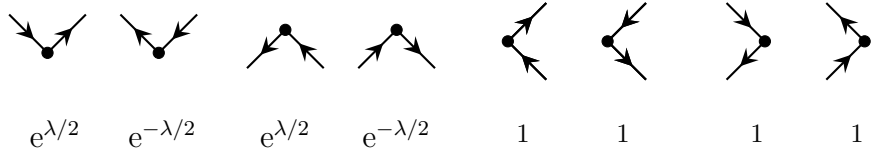

\vspace{0.6cm}


Let $T_1$ be the row-to-row transfer matrix for an odd row of $\cal L'$, and $T_2$ the 
transfer matrix for an even row. Then
\be \label{pfn6v}
Z_{6V} \eq   <\!0 \, |\,  T_1 \, T_2 \, T_1 \cdots T_2 \, T_1 | \, 0\!> \comma  \ee
where there are $M$ factors $T_1$ in the matrix product, and $M-1$ factors $T_2$, and
$ <\!0 \, |$ , $| \, 0\!> $ are vectors that account for the bottom and top boundaries of 
$\cal L'$.  Let $\Lambda^2$ be a typical eigenvalue of $T_1 T_2$, given by
the equations
\be \label{eigval} \Lambda f = T_1 g \sep \Lambda g = T_2 f  \comma \ee
$f,g$ being the associated eigenvectors.

The the right-hand side of (\ref{pfn6v}) can be written as a sum over terms, each 
proportional to  $\Lambda^{2M}$. In the limit of $M$ large, this will be given by
\be Z_{6V} \eq  C  \, \Lambda_{\rm max} ^{2M} \left[ 1+ O( \e^{-\gamma M } ) \right] 
\comma \ee
where $\Lambda_{\rm max}$ is the maximum eigenvalue and $Re(\gamma) >0 $. 
In the limit of $M$ large it follows that
\be  \label{pfntm}
\lim_{M \rightarrow \infty} \left( \log Z_{6V} \right)/M  \eq  
\log \Lambda_{\rm max} ^2 \period \ee


\section{The self-dual Potts model, with $x_1 x_2 = 1$}
\setcounter{equation}{0}

For general $x_1, x_2$ the Bethe ansatz does not work for this inhomogeneous model.
However, if  $x_2 = 1/x_1$, we can define
\be \label{xxx}
x_1 = x \sep x_2  = 1/x   \comma \ee
and then the weights for the internal  vertices on odd and even rows, 
given in (\ref{oddrows}) and  (\ref{evenrows}) satisfy 
\be  ( \omega_1, \ldots , \omega_6  )_{\rm odd} \eq x^{-1}
\;  ( \omega_1, \ldots , \omega_6  )_{\rm even}  \ee
so 
\be \label{6Vhom}
Z_{6V} \eq x^{-N(M-1)} \, Z_{\rm hom} \comma \ee
where $Z_{\rm hom}$ is the partition function of  a six-vertex model defined in the same way
as previously, but with all internal weights given by (\ref{oddrows}), so it is homogeneous 
(but not rotation-invariant).
 We  note from 
(\ref{freeenergies}), (\ref{Pottsand 6V}), (\ref{pfntm}), (\ref{6Vhom}) that
\be \label{Mlarge}
-N f_b - f_s \eq (N/2) \log Q - N \log x + \log \Lambda_0^2 \ee
to within terms of order $\e^{-\delta ' N}$, $\Lambda_0$ being the maximum eigenvalue
of the transfer matrix of the homogeneous model.

The corresponding  Potts model is self-dual, with  
\be \label{dual}  K_1^{*} = K_2 \sep  K_2^{*} = K_1 \period \ee
One must still distinguish
between $T_1$ and $T_2$ because the boundary conditions are different for the 
two type of row. However,  Owczarek and  Baxter{\color{blue}\cite{OB1989}} were able to 
solve (\ref{eigval}) by extending the Bethe ansatz to free boundary conditions 
(for every wave number $k$ there is a reflected wave number $-k$). 

The number $n$ of down arrows between two successive rows of $\cal L'$ is conserved in 
this model. Owczarek and  Baxter{\color{blue}\cite{OB1989}} solved  (\ref{eigval}) for 
arbitrary $n$, but the top and bottom boundary conditions ensure that $n=N$ (there are as 
many down arrows as up ones), and we shall only consider this case.

Our notation here is not quite consistent with {\color{blue}\cite{OB1989}}, one significant 
difference being that  $N$ in that paper is $2N$ here.

To make the notation for the weights 
consistent, associate an extra weight $t$ with the top of every down-pointing NW -SE arrow, 
and a weight  $1/t$ with the bottom of every such arrow. Then the first four weights 
$\omega_1, \ldots  , \omega_6$ in Fig. (\ref{sixvertices}) are unchanged, while
$\omega_5 , \, \omega_6$ become $t^{-1}\omega_5 ,\, t  \, \omega_6$. The eight boundary 
weights in Fig. (\ref{bdyweights}) are multiplied by $t^{-1},1,1,t, \, 1,t,t^{-1},1$, respectively.
Taking $t$ to be as in {\color{blue}\cite{OB1989}}, and $\lambda$ herein to be given by
\be \e^{\lambda/2}  \eq t \comma \ee
we obtain the weights of (2.64) -- (2.67) of  {\color{blue}\cite{OB1989}}, $q$ therein 
being the $Q$ of this paper.

These additional edge weights cancel out of the partition function and of the eigenvalue 
$\Lambda$.

 The parameter  $\mu$ of  {\color{blue}\cite{OB1989}} is given by $\mu = i \lambda $
 and we replace $ v $ therein by  $v = \mu - 2 i u$ so
 \be \label{defx}
 x = \frac{\sinh ( \lambda-2 u) }{\sinh 2 u } \period \ee
  Then 
equations (2.86), (2.87), (2.74)  of {\color{blue}\cite{OB1989}} become (replacing 
$n,N$ therein by $N, 2N $)
\be \label{eigen}
\Lambda^2 \eq \prod_{j=1}^N \frac{\sinh ( \lambda-u-\alpha_j ) \sinh ( \lambda-u+\alpha_j )}
{\sinh ( u-\alpha_j ) \sinh ( u+\alpha_j )} \comma \ee
where $\alpha_1, \ldots , \alpha_N$ are given by the $N$ ``Bethe equations"
\bd  \left[ \frac{\sinh (u+\alpha_j) \sinh (\lambda-u+\alpha_j)}{\sinh (u-\alpha_j) 
\sinh (\lambda-u-\alpha_j)} \right]^{2N}  = \spce \spce \spce \spce   \spce \spce \spce \spce  
 \spce \spce \spce \spce   \nonumber \ed
\be \label{Betheeqns}
\prod_{m=1, m \neq j}^N \frac{\sinh (\lambda + \alpha_j-\alpha_m) \,\
\sinh (\lambda + \alpha_j + \alpha_m) }
{\sinh (\lambda + \alpha_m-\alpha_j) \,\sinh (\lambda - \alpha_j - \alpha_m) }
\negspce \negspce  \ee
for $j=1, \ldots , N$. 

(\ref{Betheeqns}) has many solutions, corresponding to the various eigenvalues. We are only 
concerned with the maximum eigenvalue.


\subsection{Solution of the Bethe equations}
If the number of states $Q$ is less than four, then $\lambda$ is pure imaginary and the large-$N$
solution of (\ref{Betheeqns})  is given  in {\color{blue}\cite{OB1989}}. 

If $Q > 4$, then $\lambda$ is real and positive. For the ferromagnetic Potts model,
from (\ref{weights}) $x$ is real and positive so 
\be \label{restr}   0 < u  < \lambda /2 \period \ee

 Here we obtain the large-$N$ behaviour of the 
maximum eigenvalue  $\Lambda_{\rm max}$ for this case, using a method similar to 
that given in   Appendix D of {\color{blue}\cite{RJB1972}} for the eight-vertex model.

First write (\ref{eigen}), (\ref{Betheeqns}) in terms of polynomials in the 
variables 
\be \label{defqw} q = \e^{-2 \lambda} \sep w = e^{-2 u} \sep z_j = e^{-2 \alpha_j } \ee
as
\be \label{eigen2}
\Lambda^2 \eq  \left( w^{2N}/q^N \right) \, \prod_{j=1}^N \frac{(1-
q/w z_j ) (1- q z_j /w) }{(1- w/z_j ) (1- w z_j) } \comma \ee

\bd z_j^{-4N} \left[ \frac{(1-w z_j)( 1-q z_j/w)}{(1-w/z_j)( 1-q /w z_j)} \right)^{2N} 
\eq \spce \spce \spce \spce   \spce \spce \spce \spce  
 \spce \spce \spce \spce   \nonumber \ed
\be  \label{beeqns}
z_j^{2-2N} \frac{(1-q/z_j^2)}{(1-q z_j^2)} \prod_{m=1}^N \frac{(1-q z_j z_m)(1-q z_j/z_m)}
{(1-q z_m /z_j)(1-q /z_j z_m)}   \sep j = 1, \ldots , N \period \negspce   \ee
\vspace{5mm}

Consider the limit when $q, w \rightarrow \infty$.
From (\ref{restr}), the largest of $q, w, q/w$ is $w$, so if we take $w \rightarrow 0$,
then it is also true that  $q, q/w \rightarrow 0$. Suppose that  $z_1, \ldots, z_N $
 remain of order one. Then (\ref{beeqns}) becomes
\be \label{eqz}
  z_j^{2N+2} = 1  \sep j = 1, \ldots , N  \period \ee
This has $2N+2$ solutions for $z_j$. 

The Bethe ansatz used in   {\color{blue}\cite{OB1989}} is a sum over all permutations
and inversions of $z_1, \ldots z_N$. If any $z_j $ is equal to its inverse, or if any two 
are equal  to one another, or to their inverses, then the Bethe ansatz gives a zero 
eigenvector, which must be rejected. Replacing any $z_j$  (or $z_m$) in (\ref{eigen2}), 
(\ref{beeqns}) by its inverse does not change the equations.

We therefore reject the  solutions $z_j = \pm 1 $ of  (\ref{eqz}), and group the 
remaining $2N$ solutions into $N$ distinct pairs $z_j, 1/z_j$. Equivalently,
we require $z_1, \ldots , z_N$ to be distinct and to lie in the upper half of the 
complex plane.

Then there is a unique solution of (\ref{eqz}) for  the $z_1, \ldots , z_N$ , and the 
corresponding eigenvalue in this limit is 
\be \Lambda^2 = w^{2N}/q^N \period  \ee
This is indeed then the maximum eigenvalue $\Lambda_0$, corresponding to 
all the left-hand arrows in $\cal L'$ being 
down, and the arrows then alternating in direction from left to right. The $N$ vertices
in odd rows are in configuration 5, those in even rows in configuration 6.

Now define the functions

\be r(z) = (1-w z)^{2N} (1-q z/w)^{2N} (1-q z^2) \comma \ee
\be R(z) \eq \prod_{m=1}^N (1-z/z_m)(1-z \, z_m) \comma  \ee
\be S(z) \label{eqnS}
\eq  \frac{z^{2N+2}  \, r(1/z)} {R(q/z)} - \frac{r(z)}{R(q z)} \period \ee

Then (\ref{eigen2}), (\ref{beeqns}) can be written simply as 
\be \label{Lam}
\Lambda^2 \eq \frac{w^{2N} \, R(q/w)}{q^N \, R(w) } \comma \ee
\be  S(z_j)  \eq  0    \sep j = 1, \ldots , N  
\period \ee

$S(z)$ therefore has zeros when  $ = z_m$ or $z=1/z_m$. It also has zeros at $ z= 1 $
and $z=-1$.  It is of course a rational function, but if we take $z, z_m$ to be of order unity
and expand in powers of $q, \, w$ and $q/w$, then  to order $w^{2N}$, $S(z)$ 
remains a polynomial of degree $2N+2$.
To this order therefore, we can set
\be S(z) = (z^2-1) R(z) \period \ee
Further, the terms proportional to $z^{2N+2}, z^{2N+1}, z^{2N}, \ldots ,
z^{N+2}$ come solely from the first term on the RHS of (\ref{eqnS}), while the
terms proportional to $1, z, z^2, \ldots z^N$ come from the second term.
Using the second feature, it follows that for $|z| <1 $,
\be \label{res}
 \frac{r(z)}{R(q z)} \eq (1-z^2) R(z) \period \ee
More accurately, if $|z| < \e^{-\delta}$, then (\ref{res}) is true to 
relative order $ \e^{-N\delta}$.

Since this is true for $|z| <1$, it is more strongly true for $|z|<q$, so we can replace
$z$ by $qz$ to obtain
\be \label{res2}
\frac{r(qz)}{R(q^2 z)} \eq (1-q^2 z^2) R(qz) \period \ee
Proceeding in this way, noting that $R(z) \rightarrow 1$ as $z \rightarrow 0$,
we can solve the equations (\ref{res}), (\ref{res2}), $\ldots$ ,
for $R(z)$ to obtain
\be R(z) \eq \prod_{k=0}^{\infty} \frac{( 1 \! - \! q^{4k+2}z^2) \, r(q^{2k} z)}{\; \; (1\, - \, q^{4k}z^2)
 \; \;  r(q^{2k+1} z)} \sep |z| <1  \comma \ee
 i.e.
 \bd R(z) = \prod_{k=0}^{\infty} \frac{(1-q^{4k+1} z^2)(1-q^{4k+2} z^2)}
 {(1-q^{4k} z^2)(1-q^{4k+3} z^2)}
 \left[ \frac{(1-q^{2k} w z )(1-q^{2k+1} z/w)}{(1-q^{2k+1} w z )(1-q^{2k+2} z/w)}  \right]^{2N} \ed
 or
 \be \label{resR}
 \log  R(z) \eq \sum_{n=1}^{\infty} \frac{(1-q^n) z^{2n} }{n (1+q^{2n} ) }  - 
 2 N  \sum_{n=1}^{\infty} \frac{(w^n+q^n/w^n) z^n}{n (1+q^n ) } \period \ee
 
 \subsubsection{The free energies} 
 
 Substituting (\ref{resR}) into (\ref{Lam}), we get, to within additional terms that vanish 
 exponentially  fast as $N$ becomes large,

 \bd  \log \Lambda_0^2  =  N \!  \left[ \log \frac{w^2}{q}  +   \!  2 
 \sum_{n=1}^{\infty} \frac{(w^{2n} \!  - \! q^{2n}w^{-2n})} {n (1+q^{n})}  \right] -  
  \sum_{n=1}^{\infty} \frac{(1 \! - \! q^n)(w^{2n} \! - \! q^{2n}w^{-2n})} {n (1+q^{2n})} \ed
  so from (\ref{Mlarge}), the bulk and surface free energies of the original Potts model of 
  (\ref{Pottspartnfn}) and (\ref{freeenergies}) are
  \be \label{resfb1}
  f_b \eq  -\half \log Q + \log x - \log \frac{w^2}{q}  -   2
 \sum_{n=1}^{\infty} \frac{(w^{2n} \!  - \! q^{2n}w^{-2n})} {n (1+q^{n})}  \comma \ee
  \be  \label{surface} 
  f_s \eq   
  \sum_{n=1}^{\infty} \frac{(1 \! - \! q^n)(w^{2n} \! - \! q^{2n}w^{-2n})} {n (1+q^{2n})}  
  \period  \ee
  {From} (\ref{weights}), (\ref{defx}) \bd Q  = q+2+  q^{-1}\sep 
  x =\frac{w^2 (1-q/w^2)}{ q^{1/2} (1-w^2)}  \comma \ed  so   \be \label{resfb}
  f_b \eq \log \left( \frac{q}{1+q} \right)  - \sum_{n=1}^{\infty} \frac{(1-q^n)(w^{2n}+
  q^n/w^{2n})}{n(1+q^n)} \ee
  which is the same result as that of  eqns.  (12.5.5) and (12.5.6c) of 
  {\color{blue}\cite{book}} , $q, \psi, \beta$ therein being the $Q, f_b, \lambda-2u$ of 
  this paper. We can also write (\ref{resfb}), (\ref{surface}) as
    \be \label{resfb3}
  f_b \eq  - K_1-K_2 - \log (1+q) + \sum_{n=1}^{\infty} \frac{q^n \, (1-q^n)(w^{2n}+
  q^n/w^{2n})}{n(1+q^n)} \comma  \ee
  
   \be \label{surface2}
  f_s \eq    \log\left( \frac{1 \! - \! q^2/w^2}{1 \! - \! w^2} \right) -
  \sum_{n=1}^{\infty} \frac{q^{n} (1 \! + \! q^n)(w^{2n} \! - \! q^{2n} w^{-2n})} {n (1+q^{2n})}    \period  \ee

 Rotating  the model through $90^{\circ}$ is equivalent to inverting $x$, i.e. of 
  replacing  $u$ by $\lambda/2 -u$, and of replacing $w$ by $q^{1/2}/w$. We see that this 
  does indeed leave the RHS of (\ref{resfb}) unchanged. Also,  making this rotation
  we obtain from (\ref{surface}) the result
  \be \label{resfsp}
   f'_s \eq   
  \sum_{n=1}^{\infty} \frac{q^n (1 \! - \! q^n)( w^{-2n} \! - \! w^{2n})} {n (1+q^{2n})}  
 \ee
 for the horizontal surface free energy.
 

 \section{The isotropic case conjectures of Vernier and Jacobsen}
 \label{sec4}
 \setcounter{equation}{0}

 \subsection{Bulk and surface free energies} 

Vernier and Jacobsen{\color{blue}\cite{VJ2012}}  negated the free energies, here we 
revert to the conventional signs, as given in (\ref{freeenergies}). As we noted earlier,
if $q_{VJ}$ is their $q$, then our $q = q_{VJ}^2$.
For the rotationally 
invariant case, when $w= q^{1/4}$, they obtained 
\be 
\e^{-f_b} \eq  \frac{(1+q)}{q (1-q^{1/2})^2} \; \prod_{k=1}^{\infty} 
\left( \frac{1- q^{2k-1/2}}{1-q^{2k+1/2} }\right)^4  \period \ee
Taking logarithms, this gives
\be f_b \eq \log \left( \frac{q}{1+q} \right)   - 2 \sum_{n=1}^{\infty} \frac{q^{n/2} \, 
(1-q^{n})}{n (1+q^{n})} \period \ee
They observed that this does indeed agree with the known result (\ref{resfb}) above.

They also conjectured that
\be  \e^{-f_s} \eq (1-q^{1/2}) \prod_{k=1}^{\infty} 
\left( \frac{1-q^{4k-1/2}}{1-q^{4k-5/2}} \right)^2 \comma \ee
i.e.
\be \label{conjfs}
f_s \eq \sum_{n=1}^{\infty} \frac{q^{n/2} (1-q^n)^2}{n (1+q^{2n})} \period \ee
Again, this  agrees with the our result (\ref{surface}) when $w = q^{1/4}$.

   
 \subsection{The corner  free energy} \label{42}
 Vernier and Jacobsen{\color{blue}\cite{VJ2012}}  also conjectured from their series 
 expansions that the corner free energy is given by
  \be \label{conj}
  \e^{-f_c} \eq \prod_{k=1}^{\infty} \frac{1}{(1-q^{4k-3})(1-q^{4k-2})^4 (1-q^{4k-1})}
  \comma \ee
  i.e.
\be  \label{cnrfree} f_c \eq   -  \sum_{n=1}^{\infty}    \frac{q^n+4 \, q^{2n} + q^{3n}}{n (1-q^{4n}) }
\period \ee 

 \subsection{Our series expansions} 
 \label{43}
We have also used series expansions to test  Vernier and Jacobsen conjectures. We put 
the six-vertex model
 into interaction-round-a-face (IRF) form{\color{blue}\cite[\S 10.3]{book}} and calculated
 the finite-size partition function by dividing it into four corners, as in the corner transfer
 matrix method{\color{blue}\cite[Fig. 13.2]{book}}, and building up the lattice by going round
 the centre spin.  We took
 \be w = q^{1/4} s^{1/2} \ee
 and expanded $f_b, f_s, f'_s, f_c$ in powers of $q$ for given $s$.  The coefficients of 
 the expansion are Laurent polynomials in $s$,  and  in the
 isotropic (rotation-invariant)  case $s$ is equal to one.
 
 This was reasonably efficient, but we were only able to get to order $q^9$, 
whereas  Vernier and Jacobsen{\color{blue}\cite[\S3.2]{VJ2012}}  went to order $q^{31/2}$.
We of course agreed with them for  $s=1$.

For general $s$, we found, to the order to which we went, that $f_c$ was {\em independent} 
of $s$ (i.e. all the  coefficients were constants), suggesting that this is true to 
all orders and $f_c$  is exactly independent of $s$ or $w$, being a function only of $q$.
This agrees with our result for $f_c$ of the next section.


 \section{Inversion relations}
 \label{sec5}
 \setcounter{equation}{0}
 
 {From} (\ref{weights}) and  (\ref{xxx}),
 \be \e^{K_1} = 1 + Q^{1/2} x \sep \e^{K_2} = 1 + Q^{1/2} /x
 \comma \ee
 so from (\ref{defx}),
 \ba \label{K1K2}
  \e^{K_1}  & = &  \frac{\sinh (2 \lambda-2 u)}{\sinh 2 u }  \eq \frac{w^2}{q} \;
  \frac{1- q^2/w^2}{1-w^2} \comma \nonumber \\
  \e^{K_2}  & = &  \frac{\sinh ( \lambda+2 u)}{\sinh (\lambda -2 u )} \eq
 \frac{1}{w^2} \;
  \frac{1- q w^2}{1-q/w^2} \period \ea
  We regard these equations as defining $K_1,K_2$ as functions of
  the variable $u$. Then
  \be \e^{K_1(u) } \, \e^{K_1(\lambda-u)}  = 1 \sep
  \e^{K_2(\lambda-u) }  = \frac{ \sinh(3 \lambda-2u)}{\sinh(2u-\lambda) }
  \eq  2-Q-\e^{K_2(u)} \ee
    
  The row-to-row transfer matrix of the Potts model, as formulated in (\ref{Pottspartnfn}), is 
  $\tilde{T}_1 \tilde{T}_2$, where 
  \be \label{defT1T2}
  (\tilde{T}_1)_{{ \sigma},{\sigma}' } \eq  \delta(\sigma,\sigma')  
  \prod_{j=1}^{N-1} \e^{K_1 \delta({\sigma}_j, {\sigma}_{j+1}) }
 \sep (\tilde{T}_2)_{{ \sigma},{\sigma}' } \eq 
  \prod_{j=1}^N \e^{K_2 \delta({\sigma}_j, {\sigma}'_{j}) }
  \ee
  writing $\sigma = \sigma_1, \ldots, \sigma_N$ for all the $N$ spins in a row, and 
  similarly for  the spins $\sigma'= \sigma'_1, \ldots, \sigma'_N$ in the row above.
  Regarding $\tilde{T}_1, \tilde{T}_2$ as functions of the variable $u$, it follows that
  \be \label{invT}
   \tilde{T}_1(u)  \tilde{T}_1(\lambda -u)  = {\bf{1} }\sep
   \tilde{T}_2(u)  \tilde{T}_2(\lambda -u) \eq \xi(u)^N {\bf{1} } 
   \comma \ee
   where $\bf{1}$ is the $Q^N$-dimensional identity matrix and
   \be \xi (u) \eq \e^{K_2(u)} \e^{K_2(\lambda-u)} +Q-1 \eq - \, \frac{  Q \sinh (2u) 
   \sinh (2 \lambda-2u)}{\sinh (\lambda-2 u )^2 } \period  \ee
   
    Define the combined transfer matrix 
   \be \label{defV}
   V = T_2^{1/2} T_1 T_2^{1/2}   \ee
and let $| 0 \rangle$ and $\langle 0 |$ be the  $Q^N$-dimensional column and row  vectors 
all of whose entries are one. Then from (\ref{Pottspartnfn})
\be \label{partf}
Z_P \eq  \langle 0 | T_1 T_2 T_1 \cdots  T_2 T_1  | 0 \rangle \eq  
\langle 0 | T_2^{-1/2}  \, V^M  \, T_2^{-1/2}  | 0 \rangle  \period \ee


Let \be \Delta \eq \Delta (u) \eq  \e^{K_2} + Q -1 \eq \frac{ 2 \,  \cosh \lambda \, \sinh (2 \lambda-2u)}
{\sinh( \lambda-2 u ) } \comma \ee
then {from} (\ref{defT1T2}),  
\be T_2 \,  | 0 \rangle \eq \Delta ^N  \,  | 0 \rangle \comma  \ee
so  $  | 0 \rangle $ is an eigenvector of $T_2$ and 
\be T_2^{-1/2}  \,  | 0 \rangle \eq  \Delta^{-N/2}   \,  | 0 \rangle \period  \ee
Hence  (\ref{partf}) can be written
\be \label{partfn }
Z_P \eq  \Delta^{-N}  \,  \langle 0 |  V^M   | 0 \rangle  \period \ee

The $\Lambda^2$ of (\ref{eigval} ) is also the eigenvalue of $V$, so if we neglect only terms
that are relatively exponentially small when $M$ is large, we can write (\ref{partfn }) as
\be \label{pfn} Z_P \eq  \Delta^{-N} \,  \Lambda_{\rm max} ^{2M}  \, \langle \psi | 0 \rangle^2 
\comma \ee
where $\psi$ is the maximal eigenvector of $V$:
\be V \psi \eq  \Lambda_{\rm max} ^2 \, \psi \period \ee

The number of rows $M$ enters (\ref{pfn}) only explicitly ($\Lambda_{\rm max}$ and 
$\psi$ are independent of $M$), so from (\ref{freeenergies}),
\be \label{relf}
-N f_b - \! f_s =   2 \log \Lambda_{\rm max} \sep \!  -N f'_s - \! f_c =   -N \log  \Delta + 
2 \log \langle \psi | 0 \rangle \period \ee

We expect these equations to hold in the physical region, where 
$ 0 < u < \lambda/2 $ and all the Boltzmann weights are positive. We would like to analytically
continue them to $u > \lambda/2$.
 
For the Potts model turned through $45^{\circ}$, with cylindrical boundary conditions, this 
is not difficult. The eigenvector $\psi$ is independent of $u$, so for finite $N$ the eigenvalue 
$\Lambda_{\rm max}$ is (after removing the known poles coming from $e^{K_1}$ and 
$\e^{K_2}$) a polynomial on $w$. Here we do not have these properties, but we shall show 
that if we make some plausible analyticity assumptions, then we can obtain the results
(\ref{resfb1}) - (\ref{resfsp}) very simply.

{From} (\ref{invT}) and (\ref{defV}), exhibiting the dependence of $V$ on $u$,
\be \label{VV}
V(u) V(\lambda-u) \eq \xi(u)^N\,  \bf 1 \period \ee
Hence if $\psi$ is the maximal eigenvector of $V(u)$, it is also an eigenvector of
$V(\lambda-u)$. Let $\Lambda (u)$ and $\Lambda(\lambda-u)$
be the associated eigenvalues. (For $ 0 < u < \lambda/2 $  
the latter will be the smallest of the eigenvalues.) Then 
\be \label{lamlam}
\Lambda (u)^2 \,  \Lambda(\lambda-u)^2  \eq  \xi(u)^N\period \ee
This relation defines $\Lambda(u)$ is the larger interval $0 < u < \lambda$.
We {\em assume} that the resulting function
$\Lambda (u)$ is analytic throughout this extended interval, in particular 
at the inversion point $u = \lambda/2$ (apart from a trivial 
pole of degree $N$ coming from the double pole of $\xi (u)$).
\addtocounter{equation}{1}
\setcounter{storeeqn}{\value{equation}}
\setcounter{equation}{0}
\renewcommand{\theequation}{\arabic{section}.\arabic{storeeqn}\alph{equation}}

We also assume that the relations (\ref{relf}) can be analytically continued into
the extended interval. Then on replacing $u$ by $\lambda-u$ in the first relation
and using (\ref{lamlam}), we obtain
\be \label{518a}
-N f_b(\lambda-u) -f_s(\lambda-u) \eq N \, \log \xi(u) -2 \, \log \Lambda_{\rm max}  \ee
where $\Lambda_{\rm max} = \Lambda(u)$.
Doing the same in the second relation gives
\be \label{518b}  - N f'_s(\lambda-u) -f_c(\lambda-u) \eq -N \log \Delta(\lambda-u) + 2 \log
 \langle \psi | 0 \rangle  \comma \ee
 $\psi$ being unchanged. 
\setcounter{equation}{\value{storeeqn}}
\renewcommand{\theequation}{\arabic{section}.\arabic{equation}}

\addtocounter{equation}{1}
\setcounter{storeeqn}{\value{equation}}
\setcounter{equation}{0}
\renewcommand{\theequation}{\arabic{section}.\arabic{storeeqn}\alph{equation}}

Adding (\ref{518a}) to the first of the  relations (\ref{relf}) (exhibiting the dependence on $u$), 
we eliminate $\Lambda_{\rm max}$. Then
separating the terms linear in $N$ from those independent of $N$, we  obtain
\be \label{inv1}
- f_b (u) -f_b(\lambda-u) \eq  \log \xi (u)  \sep - f_s(u) -f_s(\lambda-u) \eq 0 \period \ee
Subtracting (\ref{518b}) from from the second relation  (\ref{relf}), we eliminate 
 $ \langle \psi | 0 \rangle $ and obtain
 \be \label{inv2}
 -f'_s(u) + f'_s(\lambda-u) \eq \log \frac {\Delta(\lambda-u) }{\Delta(u)} \sep
 -f_c(u) + f_c(\lambda-u) \eq 0 \period \ee

\setcounter{equation}{\value{storeeqn}}
\renewcommand{\theequation}{\arabic{section}.\arabic{equation}}

 We refer to the four relations (5.19) as the {\em inversion relations}. There are also four 
 {\em rotation relations} that can be obtained by noting that replacing $u$ by $\lambda/2-u$ 
 interchanges $K_1$ with $K_2$ which is equivalent to rotating the lattice through 
 $90^{\circ}$, so
 \ba \label{rotn1}
 f_b(u) = f_b(\lambda/2-u)  & , & f_s(u) = f'_s(\lambda/2-u) \!\!  \comma \nonumber \\
 f'_s(u) = f_s(\lambda/2-u)  & , & f_c(u) = f_c(\lambda/2-u) \period \ea
 

 \subsection{Alternative derivation of the free energies}
 
 We shall now show that we can use the above inversion and rotation relations 
 to derive the bulk and surface free energies, 
 and to show that the corner free energy depends only on the parameter $\lambda$, but 
 {\em not} on $u$. The method depends on certain analyticity assumptions, so is 
 not rigorous, but it is much simpler than the Bethe ansatz method used above.
 
 \subsubsection{Assumptions}

 For finite $M, N$ the partition function is a finite sum of products of $\e^{K_1}$ and 
 $\e^{K_2}$, so from (\ref{K1K2}) is a rational function of $w^2$. The denominator 
 is a product of at most $M(N-1)$ powers of  $1-w^2$, and of at most 
 $N(M-1)$ powers of $1-q/w^2$. From (\ref{freeenergies}), we therefore expect 
 $\e^{-f_b}$ to have  simple poles at $w^2=1$ and $w^2= q$, $\e^{-f_s}$
 to have a simple zero at $w^2=1$, and $\e^{-f'_s}$ to have a simple zero at
 $w^2=q$. 

 Define $F(u), G(u)$ by
 \be \e^{-f_b} \eq \e^{K_1+K_2} F(u) \sep 
 \e^{-f_s(u) } \eq \frac{(1-w^2) \, G(u)}{1-q^2/w^2}  \comma \ee
 then, consistent with the above remarks and with series expansions, we
 {\em assume} that $\log F(u), \log G(u), f_c(u) $ are  single-valued analytic
 functions of $w^2$, not just in the physical regime  $q < w^2 < 1$, but in an annulus
 containing $ q \leq |w^2| \leq 1 $ in the complex $w^2$-plane.
 
 Hence we can write
 \be \label{bexpn} \log F(u)  \eq  c_ 0^{(b)} + 
\sum_{n=1}^{\infty} [c_n^{(b)} w^{2n} + d_n^{\, (b)}  w^{-2n} ] \comma \ee
\be \label{conjG}
\log G(u) \eq  c_ 0^{(s)} + 
\sum_{n=1}^{\infty} [c_n^{(s)} w^{2n} + d_n^{\, (s)}  w^{-2n} ] \comma  \ee
\be \label{fcexpn}
f_c(u) \eq  c_ 0^{(c)} + 
\sum_{n=1}^{\infty} [c_n^{(c)} w^{2n} + d_n^{\, (c)}  w^{-2n} ] \comma \ee
where the expansions are convergent for  $ q \leq |w| \leq 1 $.

 We shall show that the relations (5.19), (\ref{rotn1}) then define the coefficients
 in these expansions, with the sole exception of $c_ 0^{(c)} $.
 This gives $f_b, f_s$ and $f_c$, and $f'_s$ is then given by
 the third of the relations  (\ref{rotn1}).

\subsubsection{Bulk free energy}

{From} (\ref{inv1}), (\ref{rotn1}) and (\ref{bexpn}),
\ba \log F_b(u)  + \log F_b(\lambda-u)  & = &    \log \xi(u) -K_1(u)-K_2(u) -K_1(\lambda - u)-
K_2(\lambda- u)  \nonumber \\
& = & 2 \log (1+q) - \sum_{n=1} \frac{(1-q^n)(w^{2n}+q^{2n}/w^{2n})}{n} \period \ea
Using (\ref{bexpn}) and equating the series term by term, this gives
\be c_0^{(b)} = -\log (1+q) \sep c_n^{(b)} + q^{-2n} \, d_n^{(b)} = (1-q^n)/n \sep n > 0 
\period \ee
Further, the first  of the  rotation relations (\ref{rotn1}) gives $\log F_b(u) =
 \log F_b(\lambda-u)$
and hence $d_n^{(b)} = q^{n} \, c_n^{(b)}$, so
\be c_n^{(b)} =   q^{-n} \, d_n^{(b)} = \frac{q^n (1-q^n)}{n(1+q^n) }  \sep n > 0 \comma  \ee
in agreement with our previous result (\ref{resfb3}).


\subsubsection{Surface free energy}
Using (\ref{rotn1}), we can write the first of the relations (\ref{inv2}) as
\be  \label{fsinv} f_s(u) - f_s(-u) \eq \log \frac{\Delta (\lambda/2-u)}{\Delta(\lambda/2+u)} 
\eq \log \left[ - \frac{\sinh (\lambda+2 u)}{\sinh (\lambda-2u)} \right]  \period \ee
Then  (\ref{inv1})  and (\ref{conjG}) give $G(u) G(\lambda-u) = 1$, and hence from (\ref{conjG})
\be c_0^{(s)} =0 \sep c_n^{(s)} +q^{-2n} d_n^{(s)} = 0 \sep  n > 0  \period \ee

Also, using (\ref{rotn1}) in the first of the relations (\ref{fsinv}), we obtain
\be f_s(u) - f_s(-u) \eq  \log \frac{\Delta(\lambda/2-u)}{\Delta(\lambda/2+u )} \eq 
\log \left[ - \frac{(1 - qw^2)}{w^2 (1-q/w^2)} \right]  \ee
which implies
\be - \log G(u)  + \log G(-u) \eq \log \frac{(1-q w^2)(1-q^2 w^2)}{(1-q /w^2)(1-q^2 /w^2) } 
 \ee
and hence, for $n > 0 $, 
\be c_n^{(s)} - d_n^{(s)} \eq \frac {q^n (1+q^n)}{n} \period \ee

It follows that
\be  c_n^{(s)} = \frac{q^n(1+q^{n}}{n(1+q^{2n})} \sep d_n^{(s)} =
 - \,  \frac{q^{3 n}(1+q^n) }{n(1+q^{2n})} \comma  \ee
 so from (\ref{conjG})
 \be \log G(u) \eq \sum_{n=1} \frac{q^n (1+q^n) (w^{2n}-q^{2n}/w^{2n})}{n (1+q^{2n})}
 \ee
in agreement with our result  (\ref{surface2}).


\subsubsection{Corner free energy}
Using (\ref{fcexpn}),   the last of the relations (\ref{inv2}), (\ref{rotn1}) give
\be d_n^{(c)} \eq q^{2n} \, c_n^{(c)} \eq q^{n} \, c_n^{(c)} \sep n > 0 \period \ee
Since $0 < q <1$, these equations imply
\be c_n^{(c)} = d_n^{(c)}  = 0 \sep n > 0  \period \ee
Hence we are left with
\be  f_c(u) \eq  c_ 0^{(c)}  \comma \ee
i.e.  $f_c(u)$ is a constant, independent of $u$, and this is in agreement with our 
conjecture of sub-section \ref{42}. If Vernier and Jacobsen's conjecture (\ref{cnrfree})
 is true for the isotropic case, 
when $u = \lambda/4$, then it follows that it must be true for all $u$.


\subsection{Inversion relations for non-solved models}

The derivations of the previous sub-section rely on $\log F(u), \log G(u)$
and $f_c(u)$ being analytic at the inversion point $u = \lambda/2, w = q^{1/2}$, 
where (\ref{VV}) implies that $V(u)$ is proportional to its inverse.
More strongly, they depend on them being analytic in a vertical strip in the 
complex $u$-plane that contains the domain $0 \leq {\rm Re} (u) \leq \lambda/2$.

There are inversion relations for models that have not been solved, e.g. the 
square lattice Ising model in a magnetic
 field,{\color{blue}\cite{RJB1980}}-{\color{blue}\cite{Maillard94}}  but the free 
 energies have complicated
singularities at the inversion point, and little progress has been made in solving them.


\subsection{Related work using the reflection relations}

Because of our assumptions regarding the analyticity properties of $\log F(u)$,
$\log G(u),  f_c(u)$, the method of this section, while simple,  is not rigorous. The 
reflection Yang-Baxter
relations{\color{blue}\cite{Pearce87}}-{\color{blue}\cite{Pearce97}}  can be used to 
obtain  functional  relations for the transfer matrix eigenvalues, and in a private 
communication{\color{blue}\cite{Pearcenotes}}  Paul Pearce shows how one can 
use these to obtain a more rigorous derivation of the inversion relations for the surface 
free energies.



 \section{Critical behaviour}
 \label{sec6}
 \setcounter{equation}{0}
 It is shown  in {\color{blue}\cite[\S 8.11]{book}}  that the bulk free energy of the six-vertex 
 model has a singularity at $\lambda = 0$, which corresponds to $Q=4$ in the Potts model.
 The singularity is of infinite order,  being proportional to $\exp (-\pi^2/\lambda)$, i.e.
 $\exp [-2 \pi^2/(Q-4)^{1/2}]$. What is the corresponding behaviour of the surface and corner
 free energies?

To answer this we need the result (4.2) of Owczarek and Baxter{\color{blue}\cite{OB1989}} 
for the surface free energy when $Q < 4$, which is  (replacing $y$  by $2y$)
\bd  f_s \eq 2\,  s_{\infty} \eq \log \frac{\sin[(\mu+v)/2]}{\sin[(\mu-v)/2]} \; \; \;  - \ed
\be \label{sinf}
 \int_{-\infty}^{\infty} \frac{2 \, \sinh (2vy) \, \sinh(\pi y -2 \mu  y) \, \cosh( \pi y -\mu y) \, 
 \cosh (\mu y) \, dy} { y \sinh (2 \pi y) \cosh(2  \mu y) } \comma \ee
where $v, \mu $ are given in terms of our $\lambda, u$ by 
\be  \label{muv}
\mu  = -i \lambda \sep v= -i (\lambda-2u)   \ee
and the $Q, x_1, x_2, x$  of ((\ref{weights}) and (\ref{xxx}) above are given by
\be Q^{1/2} = 2 \cos \mu \sep x_1 = x_2^{-1} = x =  \frac{\sin v} {\sin (\mu-v)}  \period \ee
In the physical regime(Boltzmann weights positive) $\mu, v$ are real and $ 0 <  v < \mu$.
The factor 2 in (\ref{sinf}) comes from the fact that $N' = N/2$ in (4.1) of 
{\color{blue}\cite{OB1989}}. Also, from (3.15) of {\color{blue}\cite{OB1989}}, 
$\sinh[(\pi-2\mu)y]$ in (4.2) should be $\sinh[(\pi-2\mu)y/2]$.

We can use the identity 
\bd  \sinh ( \pi y -2\, \mu y ) \cosh(\pi y - \mu y) = \sinh(\pi y -3 \mu y) \cosh (\pi y) + 
\sinh ( \mu y) \cosh( 2 \mu y) \ed
to write (\ref{sinf}) as
\bd f_s \eq \log \frac{\sin (\mu+v) }{\sin (\mu-v)} \, - \,   {\cal P} \! \! \int_{-\infty}^{\infty} 
\frac{ \sinh (2 v y) \,  \cosh (\mu y) \, \e^{ \pi y-3 \mu y}  }
{y \sinh (\pi y) \cosh ( 2 \mu y)}  \, dy \comma \ed
$\cal P$ denoting the principal value integral.

We want to analytically continue this result to $Q > 4$ so as to compare it with
(\ref{surface2}). We move $mu$ into the lower half plane and can then close the integration
round the upper-half $y$-plane. Summing the residues of the poles and suing (\ref{muv})
gives
\bd f_s =\sum_{n=1}^{\infty} 
\frac{ (1-q^n) (w^{2n}-q^{2n}/w^{2n})}{n (1+q^{2n})}   \; \; - \ed
\be  \label{anlytc}
4 \sum_{n \; \mathrm{odd} }    \frac{[ i+(-1)^{(n-1)/2} ] \, \sinh [ \pi n (\lambda -2u)/2 \lambda ] \, 
\e^{-\pi^2 n /2 \lambda } }{n (1-\e^{-\pi^2 n /2 \lambda )} } 
\comma \ee
the second sum being over all positive odd integers $n$, i.e. $n= 1,3,5, \ldots $.

Comparing this with (\ref{surface} ) above, we see that the dominant singularity  in $f_s$ is
proportional to $\e^{-\pi^2/ 2 \lambda}$. This is of infinite order, i.e. all derivatives exist and 
are continuous. This singularity is proportional to the square root of the dominant singularity
in $f_b$.

The conjectured expression (\ref{conj}) for the corner free energy can be written
\be \e^{-f_c}  \eq  P(q)^{-1}  \, P(q^2)^{-4}  \comma \ee
where
\be P(q) = \prod_{k=1}^{\infty} (1-q^{2k-1}) \period \ee
The function
\be {\cal Q}(q) \eq \prod_{n=1}^{\infty} (1-q^n) \ee
occurs in Jacobi elliptic functions and satisfies the ``conjugate modulus" relation
\be {\cal Q}(q) \eq \epsilon^{-1/2} \, \exp \left[ {\frac{\pi (\epsilon- \epsilon^{-1})}{12}} 
\right] \,  {\cal Q}(q') \comma \ee
where if $q = \e^{-2 \pi \epsilon}$, then $ q' =  \e^{-2 \pi/ \epsilon}$. Noting that
$ P(q) = Q(q)/Q(q^2)$, it follows that
\be P(q) \eq \sqrt  2  \, \exp \left[ - \frac{\pi \epsilon}{12} - \frac{\pi }{24 \epsilon}\right] 
 \; P({q'}^{1/2} ) \comma \ee
 and hence that
 \be \e^{-f_c} \eq \exp \left( \frac{ 3 \pi \epsilon}{4} +\frac{ \pi}{8 \epsilon } \right) \left/ \left[ 
 {2^{5/2} \, P({q'}^{1/2}) \, P({q'}^{1/4})^4 } \right]  \right. \ee
 in agreement with eqn. 81 of {\color{blue}\cite{VJ2012}}  (the $q$ therein is our 
 $\e^{-\pi \epsilon}$).
 
Near the critical point $Q \rightarrow 4^{+}$ and  $\epsilon, q' \rightarrow 0^{+}$. We see that
 \be f_c \sim \, -  \, \frac{\pi}{8 \epsilon} \sim \, -  \, \frac{\pi^2}{4 [2(Q-4)]^{1/2}}   \comma \ee
 so $f_c$ becomes negatively infinite.


 \section{Summary}
 \setcounter{equation}{0}
 In sections 2 and 3 we have  adapted previous work{\color{blue}\cite{OB1989}}  on the 
 $Q$-state self-dual  Potts model  on the square lattice from the case when $Q<4$ to
  when  $Q>4$. This gives the bulk free energy, which was 
  known{\color{blue}\cite[eqn. 12.5.6]{book}}, and also the vertical free energy. We 
  considered the general model, homogeneous but anisotropic. It contains two free 
  parameters, the vertical and horizontal interaction coefficients $K_1, K_2$, or 
  equivalently the parameters $q, w$ defined by (\ref{weights}), (\ref{defx}), (\ref{defqw} ).


  Vernier and Jacobsen{\color{blue}\cite{VJ2012}} had conjectured the bulk, surface 
  and corner free energies for the isotropic case, when $K_1=K_2$ and $w= q^{1/2}$.  
  We report these conjectures in section \ref{sec4}, and note that our
  results for the bulk and surface free energies, specialized to this case, agree with 
  their conjectures.  We also made  series expansions for the more general anisotropic case 
  (taking $w = q^{1/4} s^{1/2}$, where $s$ is a parameter of order unity) and found that the 
 coefficients  of the terms in the series were independent of $s$. They agreed with  
 Vernier and Jacobsen's conjectures, not just for $s=1$, but for {\em all} $s$.
    
  It is known that the bulk free energy can be easily obtained using the ``inversion
  relation" method{\color{blue}\cite{Strog}},  {\color{blue}\cite{Bax82}}, 
  {\color{blue}\cite[\S 12.5]{book}}. In section \ref{sec5} we show how this can be 
  extended to the surface and corner free energies. Together with the simple
  rotation relations and appropriate analyticity assumptions, these give an alternative
  method (simpler than the Bethe ansatz calculation of Owczarek and 
  Baxter{\color{blue}\cite{OB1989}})
  of deriving the surface free energy. They also imply that  the corner free energy
  is a function only of the number of states $Q$, in agreement  with our series expansions 
  of section \ref{sec4}.
  
  These inversion relation calculations are similar to those for the Ising 
  model.{\color{blue}\cite{Ising}}   
  
Finally, in section \ref{sec6} we discuss the behaviour when $Q \rightarrow 4^{+}$
and $q \rightarrow 1^{-}$, which is the critical case of the associated
six-vertex model.






\end{document}